\documentclass[preprint,review,12pt]{elsarticle}

\usepackage{amssymb}
\usepackage{lineno}
\usepackage{amsmath}

\usepackage{amsmath}
\usepackage{algorithm}
\usepackage[noend]{algpseudocode}

\usepackage{subcaption}

\usepackage{hhline}

\usepackage{hyperref}

\usepackage{multirow}

\usepackage{booktabs}

\journal{Expert Systems with Applications}

\begin{document}

\begin{frontmatter}

\title{A New Machine Learning Dataset of Bulldog Nostril Images for Stenosis Degree Classification}

\address[label1]{Dom Bosco Catholic University, Campo Grande, Brazil}
\address[label2]{Federal University of Mato Grosso do Sul, Campo Grande, Brazil}
%\address[label3]{Universidade Estadual de Mato Grosso do Sul, Campo Grande, Brazil}
\address[label4]{Veterinary Dentistry Clinic - OdontoPet, Campo Grande, Brazil}

\author[label1]{Gabriel Toshio Hirokawa Higa\corref{cor1}}
\ead{gabrieltoshio03@gmail.com}
\cortext[cor1]{Corresponding author}

\author[label1,label4]{Joyce Katiuccia Medeiros Ramos Carvalho}
\ead{joyce@ucdb.br}

\author[label1]{Paolo Brito Pascoalini Zanoni}
\ead{paolobrito22@gmail.com}

\author[label1]{Gisele Braziliano de Andrade}
\ead{gisele@ucdb.br}

\author[label1,label2]{Hemerson Pistori}
\ead{pistori@ucdb.br}

\begin{abstract}
%\begin{linenumbers}
Brachycephaly, a conformation trait in some dog breeds, causes BOAS, a respiratory disorder that affects the health and welfare of the dogs with various symptoms. In this paper, a new annotated dataset composed of 190 images of bulldogs' nostrils is presented. Three degrees of stenosis are approximately equally represented in the dataset: mild, moderate and severe stenosis. The dataset also comprises a small quantity of non stenotic nostril images. To the best of our knowledge, this is the first image dataset addressing this problem. Furthermore, deep learning is investigated as an alternative to automatically infer stenosis degree using nostril images. In this work, several neural networks were tested: ResNet50, MobileNetV3, DenseNet201, SwinV2 and MaxViT. For this evaluation, the problem was modeled in two different ways: first, as a three-class classification problem (mild or open, moderate, and severe); second, as a binary classification problem, with severe stenosis as target. For the multiclass classification, a maximum median f-score of 53.77\% was achieved by the MobileNetV3. For binary classification, a maximum median f-score of 72.08\% has been reached by ResNet50, indicating that the problem is challenging but possibly tractable.
%\end{linenumbers}
\end{abstract}

% The first keyword should be selected from the list of EJOR Keywords.
% Please include up to 4 additional keywords of your choice.

\begin{keyword}
computer vision \sep deep learning \sep brachycephalic dogs \sep airway obstruction

\end{keyword}

\end{frontmatter}

%\linenumbers

%%%%%%%%%%%%%% I N T R O D U C A O %%%%%%%%%%%%%%%%%%%%%%%%%%%%%%%%%%%%%%%%%%%%%%%%%%%%%%%%%%%%%
\section{Introduction}
\label{intro}

Brachycephalic dog breeds, such as the French Bulldog, have been increasing in popularity due to their flat-faced appearance, their infantile and anthropomorphic aesthetic features that owners may find endearing, including their small stature, and their low levels of aggression, which makes them popular family pets in urban settings~\cite{liu2017conformational,oneill2018demography,roedler2013how_does}. However, selective breeding for severe brachycephaly has given rise to a conformation-related respiratory disorder, known as Brachycephalic Obstructive Airway Syndrome (BOAS), which constitutes a major health and welfare problem, as it causes breathing difficulties, heat and exercise intolerance, sleep-disordered breathing, cyanosis and collapse in the affected individuals, among other clinical abnormalities of the gastrointestinal tract~\cite{liu2017conformational,luciani2022evaluation}. BOAS is a pathophysiological disorder caused by excessive soft tissue within the upper airways of short-nosed dog breeds, causing stenotic nares, aberrant nasopharyngeal turbinates, an elongated and thickened soft palate, macroglossia, and, in some cases, a hypoplastic trachea, resulting in severe respiratory distress~\cite{dupre2016brachycephalic,liu2016whole_body}. Stenotic nares are characterized by narrowed external nostrils and reduced diameter of nasal vestibules that lead into the nasal cavity. Regarding its degree, stenosis can be classified as mild, moderate, or severe~\cite{ekenstedt2020canine,liu2017conformational}. The identification of severe stenosis is a major factor in improving the well-being of the pets, but it is not a simple one, specially because the differences between degrees can be very subtle at times. In this work, therefore, we propose the utilization of Deep Neural Networks (DNNs) for the classification of stenosis degree in images of dog nostrils. The proper realization of this task shall be a huge help for tutors who care about the well-being of their brachycephalic dogs, and also for veterinarians who must deal with such cases.

In the last few years, artificial intelligence and deep learning have become a hot topic in fields that go beyond Computer Science. Its potential has already become clear in Veterinary Medicine, its domestic animals oriented branch included. In this domain, by searching through the recent literature it is possible to see that it has been mainly applied to two kinds of tasks: first, as a tool to analyze behavior in order to identify unapparent disorders and improve animal well-being, such as in the work of \citet{atif2023behavior}, who proposed an end-to-end system to identify frame sequences of dogs alone and to summarize their behavior, in order to provide useful information about abnormal behaviors to the pets' tutors; and second, and closest to our case, as a tool to analyze examination results, in order to provide assistance in the diagnosis of diseases and other physical disorders.

Among the works aimed at examinations and disease diagnostics, some works achieve good results by using shallow learning algorithms. One reason for preferring this kind of algorithm is the easiness of acquiring features organized in tabular format, such as the age of the animals or the presence or absence of a given symptom, in the occasion of a first visit to a clinic. \citet{ferreira2022diagnostic}, for instance, evaluated shallow machine learning techniques to diagnose visceral leishmaniosis in dogs. In their work, the authors tested four algorithms: Support Vector Machine (SVM), K-Nearest Neighbors, Naïve Bayes and Logistic Regression. These were used to model up to 17 features (filtered differently for each algorithm). The ELISA sorological test was taken as the dependent variable. The best results therein reported were achieved by the Logistic Regression: accuracy of 75\% and recall of 84\%. \citet{schofield2021machine} also used shallow algorithms, but to predict Cushing's syndrome. In their case, 21 features were selected and analyzed with four different models: Least Absolute Shrinkage and Selection Operator (LASSO), Random Forest, SVM and Radial Basis Function kernel SVM. The authors report that LASSO achieved the highest results: accuracy of 77\% and recall of 71\%.

When deep learning and computer vision techniques are concerned, many of the recent works have been focusing on the analysis of specialized images, such as radiographs. According to \citet{banzato2021automatic}, one of the main reasons is that the task of interpreting these examinations to settle on a diagnostic is difficult and highly error-prone. In their work, \citet{banzato2021automatic} studied the application of a ResNet50 and an EfficientNet121, two Convolutional Neural Network (CNN) architectures, to automate the classification of canine thoracic radiographs, according to the presence of abnormal conditions, such as cardiomegaly and megaoesophagus, reporting highly varying results for different conditions. Focusing more specifically on cardiomegaly, \citet{jeong2022an_automated} used a CNN to segment both the heart and the fourth thoracic vertebral body (T4), in order to calculate a novel index based on the area and height of the heart and in the length of the T4, also proposed by them, related to the presence of the aforementioned condition. Finally, \citet{baydan2021determining} proposed a modified version of a Mask R-CNN to localize fractures in radiograph images of fractured tibias of cats and dogs.

These works focus on examinations of internal abnormal conditions, such as bone fractures and cardiomegaly. There are also works that focus on external abnormalities, which is also our case. \citet{hwang2022classification}, for instance, proposed the use of CNNs to classify three different dog skin diseases: bacterial dermatosis, fungal infection, and hypersensitivity allergic dermatosis. In their case, the authors used normal images taken with a smartphone and also hyperspectral images taken with a specialized device. Dog nare pictures were used in at least one work in the last few years, the one by \citet{bae2021dog}. However, the proposal of the authors was not to use these pictures for the diagnostic of abnormal conditions, but rather to use the nose print for biometrics. The authors proposed a siamese network baptized as dog nose network, which uses pictures taken with smartphones to identify individual dogs. Biometric techniques, as the authors claim, are a viable alternative to substitute more invasive and painful identification methods, such as microchips and tags. For identification, the authors report an average accuracy of 98.972\%.

\begin{figure*}
     \centering

     \begin{subfigure}[!htb]{0.45\textwidth}
         \centering
         \includegraphics[width=0.45\textwidth,height=0.45\textwidth]{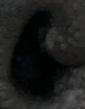}
         \includegraphics[width=0.45\textwidth,height=0.45\textwidth]{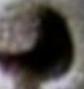}
         \caption{Open}
         \label{fig:ex_open}
     \end{subfigure}
     \hfill
     \begin{subfigure}[!htb]{0.45\textwidth}
         \centering
         \includegraphics[width=0.45\textwidth,height=0.45\textwidth]{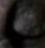}
         \includegraphics[width=0.45\textwidth,height=0.45\textwidth]{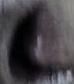}
         \caption{Mild}
         \label{fig:ex_mild}
     \end{subfigure}

     \begin{subfigure}[!htb]{0.45\textwidth}
         \centering
         \includegraphics[width=0.45\textwidth,height=0.45\textwidth]{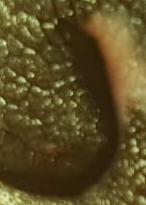}
         \includegraphics[width=0.45\textwidth,height=0.45\textwidth]{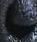}
         \caption{Moderate}
         \label{fig:ex_moderate}
     \end{subfigure}
     \hfill
     \begin{subfigure}[!htb]{0.45\textwidth}
         \centering
         \includegraphics[width=0.45\textwidth,height=0.45\textwidth]{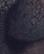}
         \includegraphics[width=0.45\textwidth,height=0.45\textwidth]{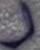}
         \caption{Severe}
         \label{fig:ex_severe}
     \end{subfigure}
     
        \caption{Samples of the collected dataset. Each image is shown with the respective stenosis degree. As discussed in Section~\ref{dataset}, since there are only three images of non stenotic (open) nares, all of them were considered images of mild stenosis.}
        \label{fig:examples}
\end{figure*}

In this work, we present a new image dataset consisting of pictures of brachycephalic dog nostrils in three different degrees: mild, moderate and severe. To the best of our knowledge, this is the first dataset addressing this problem. Figure~\ref{fig:examples} shows six examples of images pertaining to each of these three classes, along with two images of non stenotic nostrils. The importance of this problem is clear from the existing literature on this problem, and many of the justifications given system aimed at other diseases also apply, such as the necessity of improving diagnostic accuracy and that of reducing costs. Furthermore, Figure~\ref{fig:examples} shows that the creation of an artificial intelligence model for this problem is not a simple task. Although many of the social justifications for the development of assistive systems for the diagnostic of diseases in domestic animals also apply to the problem hereby proposed, many of the challenges found in this work do not. Truly, differentiating images of normal and abnormal radiographs, for instance, may require fine-grained feature extraction. In this sense, most of the works we referred to, as well as ours, could be considered difficult cases of fine-grained image classification. In our case, however, the model is also supposed to deal with images taken in a huge variety of ways and positions, since the environment in which they shall be taken is much less controlled than, \textit{e.g.}, that of a radiography room. This also implies that the system shall deal with images of different resolutions. On top of that, one can also claim that radiographs present a greater regularity than nostril pictures. Given the complexity of the proposed problem, we evaluate five neural networks that differ in novelty, size and theoretical basis. Three of these neural networks are CNNs: ResNet50, EfficientNet201 and MobileNetV3. One of them is a transformer based neural network: SwinV2. Finally, one of them is a hybrid architecture, that use both convolutions and transformer operations: MaxViT.

%%%%%%%%%%%%%%  M A T E R I A I S  E   M E T O D O S   %%%%%%%%%%%%%%%%%%%%%%%%%%%%%%%%%%%%%%%

\section{Materials and Methods}
\label{methods}

\subsection{Dataset}
\label{dataset}

This project was approved by the Ethics Committee on the Use of Animals at the Dom Bosco Catholic University, under the protocol number 010/2022. The images were taken from July 2021 to April 2023, in partnership with the OdontoPet clinic in Campo Grande-MS. The images comprise 95 brachycephalic French Bulldogs of both sexes, all over 1 year old, averaging 3.65 years, standard deviation of 2.54 years, with different degrees of body score, with stenosis of nostril. The animals were selected regardless of the presence of other anatomical alterations common to the syndrome. The animals were included after an informed consent form was signed by their guardians.

\begin{figure}
     \centering
     \begin{subfigure}[b]{0.3\textwidth}
         \centering
         \includegraphics[width=\textwidth]{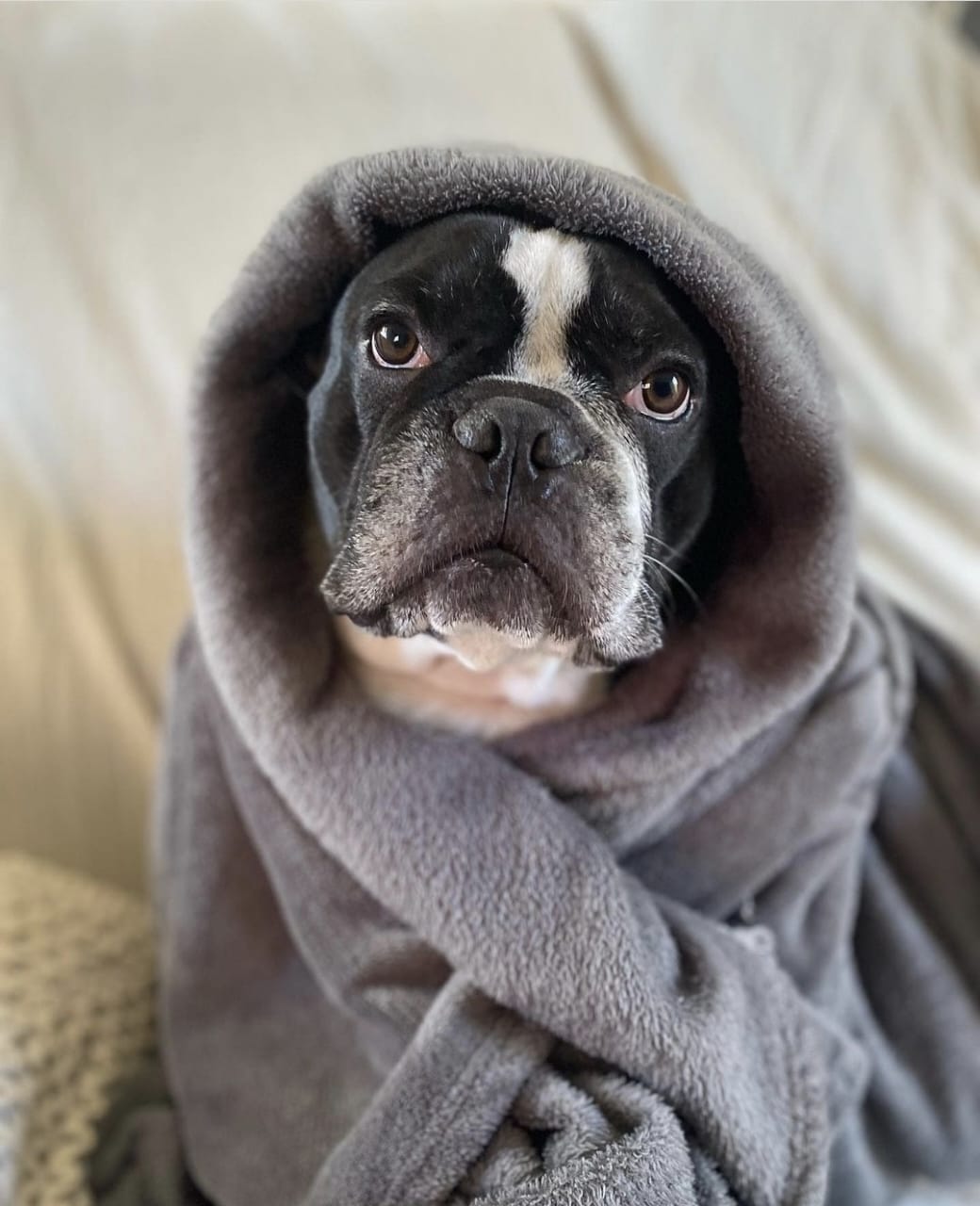}
         \caption{}
         \label{fig:original_ex_1}
     \end{subfigure}
     \hfill
     \begin{subfigure}[b]{0.3\textwidth}
         \centering
         \includegraphics[width=\textwidth]{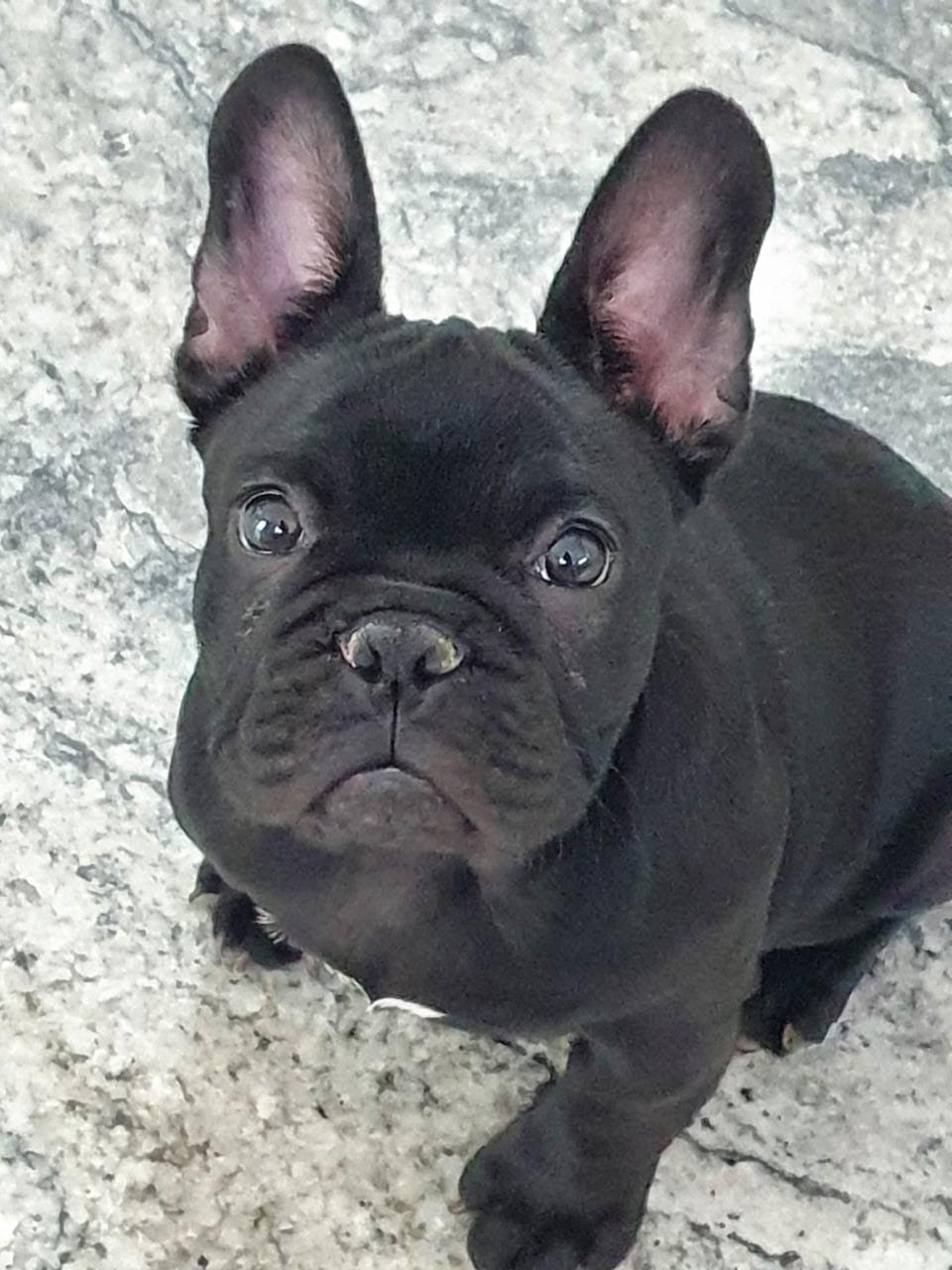}
         \caption{}
         \label{fig:original_ex_2}
     \end{subfigure}
        \caption{Two samples of the original images taken, before they were annotated and cropped for the dataset.}
        \label{fig:original_samples}
\end{figure}

Frontal photographic records of the face of each animal were taken in natural lighting conditions. The images were captured from a distance of approximately 40 cm from the dog's face, using different smartphone models with Android and IOS operating system. Direct inspection of the nostrils was carried out in order to verify the presence of stenosis. This involved not only capturing images in parallel to the face, but also with a slight tilt to the animals' faces or with the dog's tongue between the face and snout. The objective was to improve the training and performance of the neural network by including images captured in different situations. Figure~\ref{fig:original_samples} shows samples of the images taken, before they were annotated and cropped. The stenosis degree of each nare was classified by a veterinarian. Then, the images were cropped with the LabelMe tool and separated according to their classes. Samples taken from the final dataset are shown in Figure~\ref{fig:examples}.

The nostrils were assessed individually and classified as open (\textit{i.e.}, non stenotic) (A), mild stenosis (B), moderate stenosis (C) and severe stenosis (D), following the degrees of stenosis by \citet{liu2017conformational}. A prospective questionnaire was adapted from \citet{pohl2016how}, where each item was followed by a four-point Likert scale: (1) absent, (2) rarely, (3) frequent, (4) very frequent. In the questionnaire, objective questions were formulated regarding the tutor's perception of the occurrence and frequency of clinical signs that are related to brachycephalic respiratory syndrome.

The dataset created and hereby presented is publicly available.

\begin{table}
\centering
\caption{Number of images per class, both in absolute values and in percentage of the total.}
\label{table:num_images}
\begin{tabular}{|c|c|} 
    \hline
    Class & Number of images \\
    \hline\hline
    Open  & 3 (1.58\%) \\ 
    \hline
    Mild  & 56 (29.47\%) \\
    \hline
    Moderate  & 66 (34.74\%) \\
    \hline
    Severe  & 65 (34.21\%) \\
    \hline\hline
    Total & 190 \\ 
    \hline
\end{tabular}
\end{table}

\subsection{Deep Learning}

Five neural networks were evaluted: ResNet50, MobileNetV3, DenseNet201, SwinV2 and MaxViT. Along with these five neural network architectures, we evaluated the application of Sharpness-Aware Minimization (SAM)~\cite{foret2021sharpnessaware} on two optimizers: Stochastic Gradient Descent (SGD) and Adaptive Gradient (Adagrad). Next, we comment on the reasons for choosing each one of these architectures\footnote{In this work, we used the PyTorch~\cite{paszke2019pytorch} framework to implement the necessary workflows. We also used the available CNNs. For the transformer-based architectures, we used the implementation available in the PyTorch Image Models (timm) library~\cite{rw2019timm}. The SAM optimization technique utilized by us is not the official one (since there is no official PyTorch implementation of it), but the one by the GitHub user davda54, available here: \url{https://github.com/davda54/sam}}.

We begin with the three fully convolutional neural networks. CNNs have been considered to constitute the State-of-the-Art in CV ever since the AlexNet~\cite{krizhevsky2012imagenet} managed to win the LSVRC competition in 2012.

\begin{itemize}
    \item \textbf{ResNet50}: the family of Residual Neural Networks (ResNets) was proposed in 2015 by \citet{he2016deep} and can be considered a very succesful architecture. The main idea behind residual neural networks was the introduction of skip connections as a way to avoid gradient vanishing, thus allowing neural networks to get deeper. In this work we evaluated the ResNet50, the version composed of 50 layers used not only for image classification, but also as the backbone for neural networks applied to other computer vision tasks, such as object detection. In this work, the ResNet50 was chosen both for being widely used (meaning that this investigation could also give insights that go beyond the scope of this work), and also as a comparison object to help in measuring the performance of the other architectures.
    
    \item \textbf{MobileNetV3}: the history of the MobileNets goes back to 2017, when the first version was proposed by \citet{howard2017mobilenets}. The second~\cite{sandler2018mobilenetv2} and third~\cite{howard2019searching} versions proposed improvements aimed at making them more efficient. ``Efficiency'' can be considered the keyword for this family of architectures, since they aim at keeping high performance even when run on smaller devices. This characteristic was also the reason for us choosing to evaluate it in this work, since achieving high performance in the task at hand may be necessary for a future, end-to-end, system.
    
    \item \textbf{DenseNet201}: the Densely Connected Convolutional Networks (DenseNets) were proposed in 2017 by \citet{huang2017densely}. As the ResNets, this family of neural networks also aimed at solving the problem of vanishing gradients, by using similar principles. Instead of using skip connections, the DenseNets utilize dense connections, where each layer is connect to all subsequent layers within a dense block. This means that DenseNets reuse features, and have the potential of achieving higher performance than the ResNets while being more efficient. In this work, we chose to evaluate the DenseNet201, since it is similar to the ResNet50 in size (the former has around 20 million parameters, while the latter is only a bit larger, with 25 million parameters).

\end{itemize}

The three neural networks above are purely convolutional neural networks. Below, we comment on those that are based on transformers. Transformers were proposed in 2017 in the context of Natural Language Processing (NLP)~\cite{vaswani2017attention}. It became widely researched in CV specially after 2021, when \citet{dosovitskiy2021an_image} proposed the Vision Transformer (ViT) as a purely transformer-based architecture. It is well recognized that transformer-based computer vision networks have problems related to the fact that they do not deal with the data in the same way as CNNs, and also to the fact that the properties of image data are different from the properties of texts. One of these problems is the necessity of larger quantities of data. The following neural networks attempted to improve on these problems, and were chosen by us as instances of State-of-the-Art transformers whose improvements may be considered necessary to learn the dataset proposed by us.

\begin{itemize}
    \item \textbf{SwinV2}: the use of patch merging and shifted windows (as opposed to sliding, specially) for self-attention computation was proposed in 2021 by \citet{liu2021swin}. This first version was baptized Swin Transformer. Its second version, the SwinV2 architecture, was proposed in 2022 by \citet{liu2022swinv2}, and introduced a new cosine attention, a new position bias and also a residual-post-norm method.
    
    \item \textbf{MaxViT}: the MaxViT is a hybrid architecture, making use of both transformer techniques and convolutions. It was proposed by \citet{tu2022maxvit} in 2022. Similarly to the creators of SwinV2, the authors also proposed a new way to compute self-attention: in a block, for local attention, and by taking the feature map as a grid, for global attention.
\end{itemize}

\subsection{Experimental Design}

%Parágrafo 7: Dizer quais foram as métricas utilizadas, como foram feitas as repetições (E.g.: validação cruzada estratificada em 10 dobras), percentuais de treino, validação e teste, estatísticas extraídas, gráficos produzidos e os testes de hipótese aplicados. 

The images of the dataset described in Section~\ref{dataset} were organized in two different ways. First, they were separated into three classes, according to the stenosis degree: mild, moderate and severe. Since they are too few in number, the three images of non stenotic nares were put in the mild class. Going further, we will refer to this arrangement as the multiclass classification problem. Second, the images were also separated in two classes: severe and non-severe, where the latter comprises all the images of non stenotic, mildly stenotic and moderately stenotic nares. Likewise, we will refer to this arrangement as the binary classification problem. Conceptually, therefore, one can think of the first problem as that of classifying stenosis degree, and of the second one as that of identifying severe stenosis (which is the most important class for this case, since it is the one that involves the greatest suffering). Then, the neural networks were evaluated with the same hyperparameters for each one of these two problems.

For the evaluation of the neural networks, a ten-fold cross-validation strategy was used. In each fold, 20\% of the images allocated for training were used for validation. The images were resized to (256, 256), since this is required by some of the neural networks, in virtue of their implementation. The pixel values of the images were in the $[0, 1]$ range. The experiment was performed with a Nvidia RTX 3060 GPU of 12 GiB. For this reason, the batch size was set to 8, since the memory of the GPU is not sufficient to handle larger batches when training the larger networks. For the optimization, a learning rate of 0.001 was used, since this is a common practice. The maximum number of epochs was set to 1000, but validation loss values were monitored for early stopping, with a large patience of 300. A tolerance against improvements smaller than 0.1 was also used. The large patience value was chosen to provide insights on the evolution of the model during training, which could prove helpful, given the complexity of the problem. This patience notwithstanding, the evaluation on the test set in each fold was performed not with the weights of the last executed epoch, but with those that achieved the best validation loss value (tolerance considered).

Data augmentation techniques were heavily used, specially those related to the colors. There are two reasons for this. The first one is the size of the dataset, as shown in Section~\ref{dataset}. The second reason is that since the visual appearance of the stenosis degree is that of too much tissue and a small nostril, and the nostril usually appears in a darker color (usually downright black), it makes intuitive sense that applying modifications to the colors can teach the network to look at the difference between the nostril and the tissue responsible for the stenosis. Specifically, in our experiment, the following data augmentation techniques were used: color jitter, random grayscale, random invert, random solarize with a threshold of 0.75, random auto contrast, random crop, random horizontal and vertical flips with probability of 50\%, random rotation in 90\textdegree, random perspective and random adjust sharpness with a factor of 2 and probability of 50\%. These techniques were not applied to the test set.

Three metrics were calculated in each fold and used in the analysis of performance: precision, recall and f-score. For the binary classification, the severe class was taken as the positive one, since it is whose identification is the most important. For the multi-class classification, the macro averages of the metrics were calculated. Lastly, the averages, standard deviations, medians and interquartile ranges of the results across the ten folds were calculated and a two-way analysis of variance (ANOVA) was conducted with a significance threshold of 5\%, followed by the Scott-Knott clustering test, which was applied regardless of the ANOVA results. Whenever a result was found to be undefined (e.g., for implying a division by zero), its value was set to zero. Other tools generated and utilized in the analysis include boxplots, confusion matrices and gradient-weighted class activation mappings (GradCAMs).

%%%%%%%%%%%%%%  R E S U L T A D O S    %%%%%%%%%%%%%%%%%%%%%%%%%%%%%%%%%%%%%%%

\section{Results}
\label{results}
\subsection{Three classes}
\label{results_3c}

\begin{figure*}[htb!]
     \centering
     \begin{subfigure}[!t]{0.45\textwidth}
         \centering
         \includegraphics[width=\textwidth]{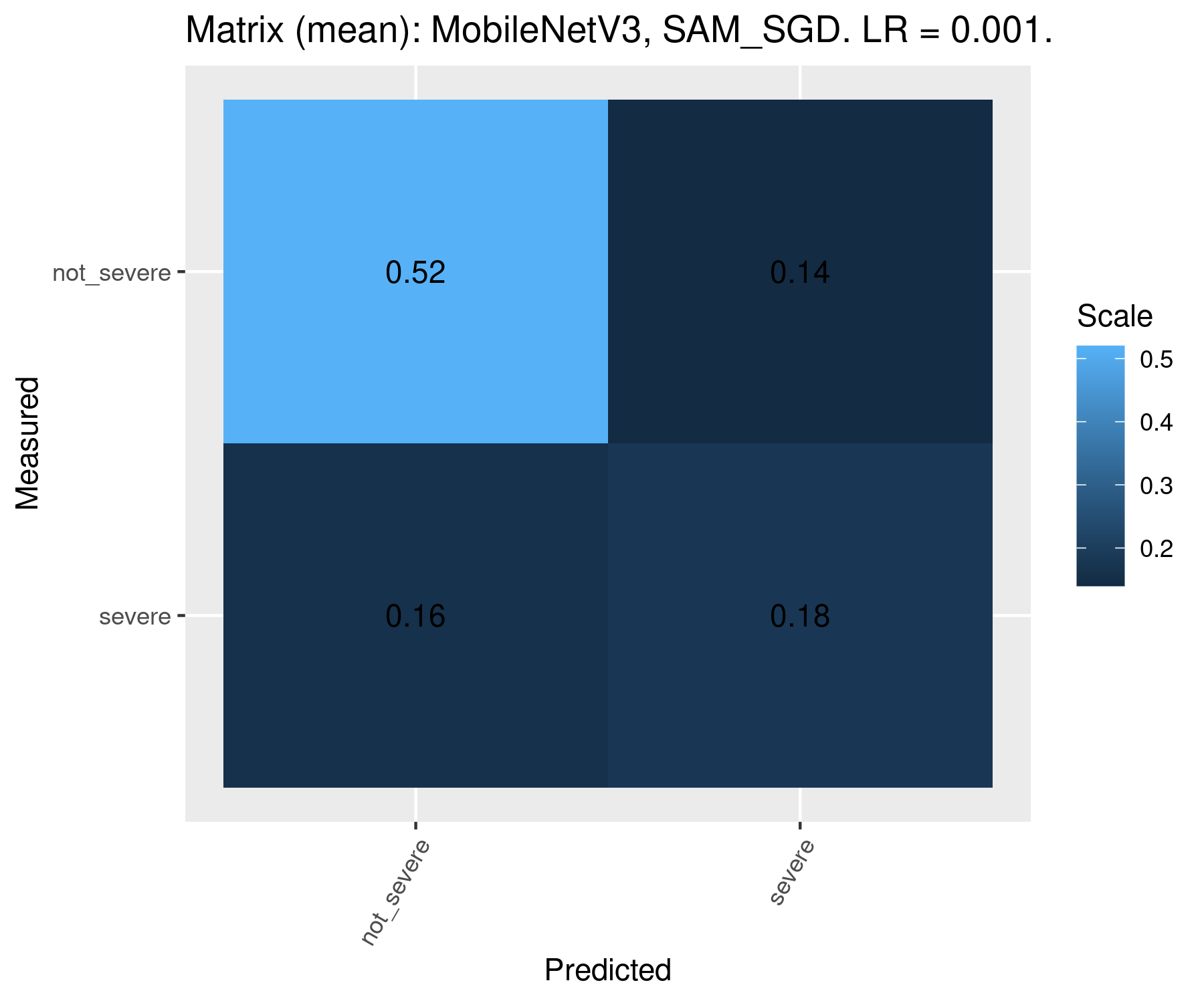}
         \caption{}
         \label{fig:cm_mobilenet_sam_sgd_2c}
     \end{subfigure}
     \begin{subfigure}[!t]{0.45\textwidth}
         \centering
         \includegraphics[width=\textwidth]{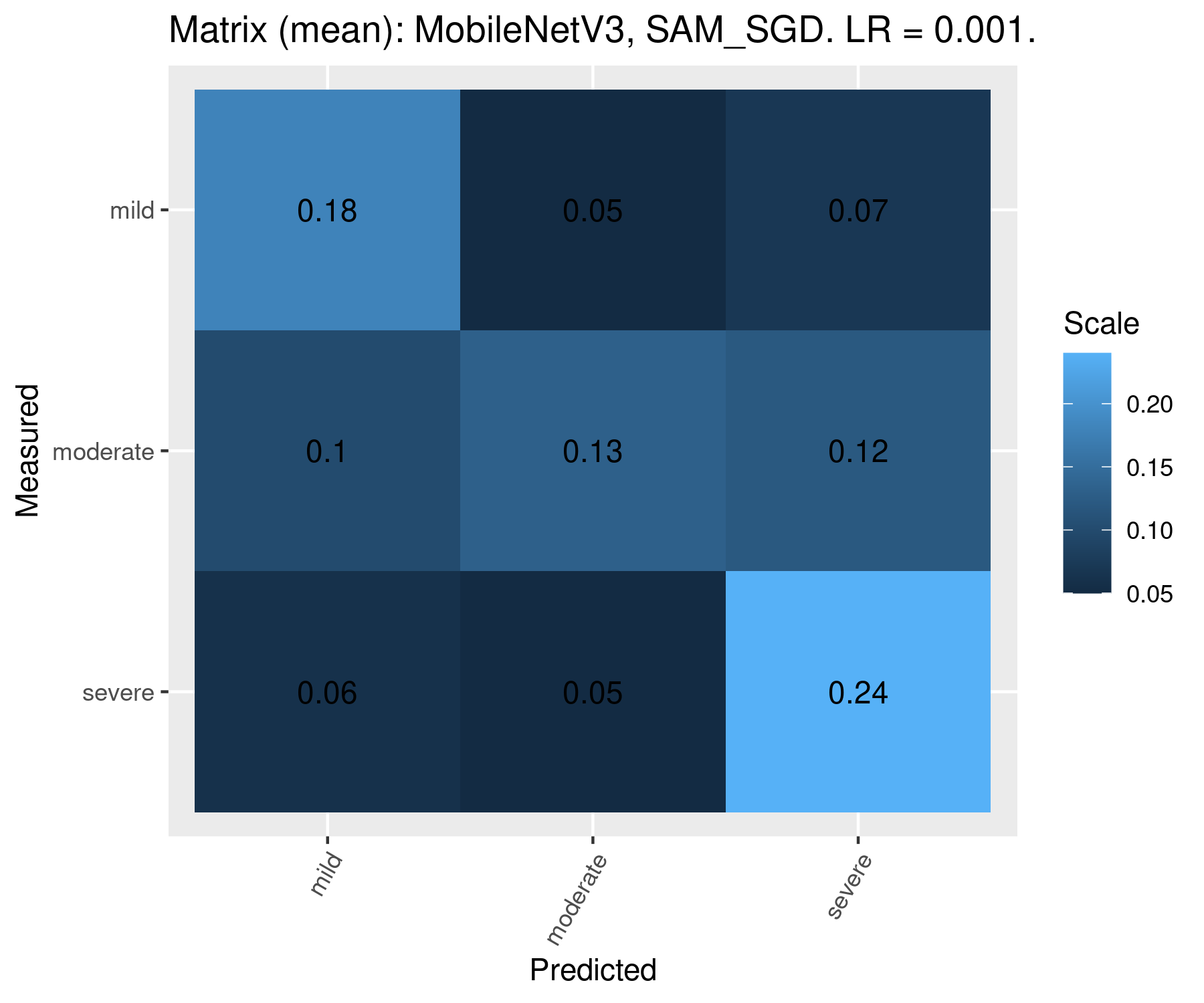}
         \caption{}
         \label{fig:cm_mobilenet_sam_sgd_3c}
     \end{subfigure}
     \begin{subfigure}[!t]{0.45\textwidth}
         \centering
         \includegraphics[width=\textwidth]{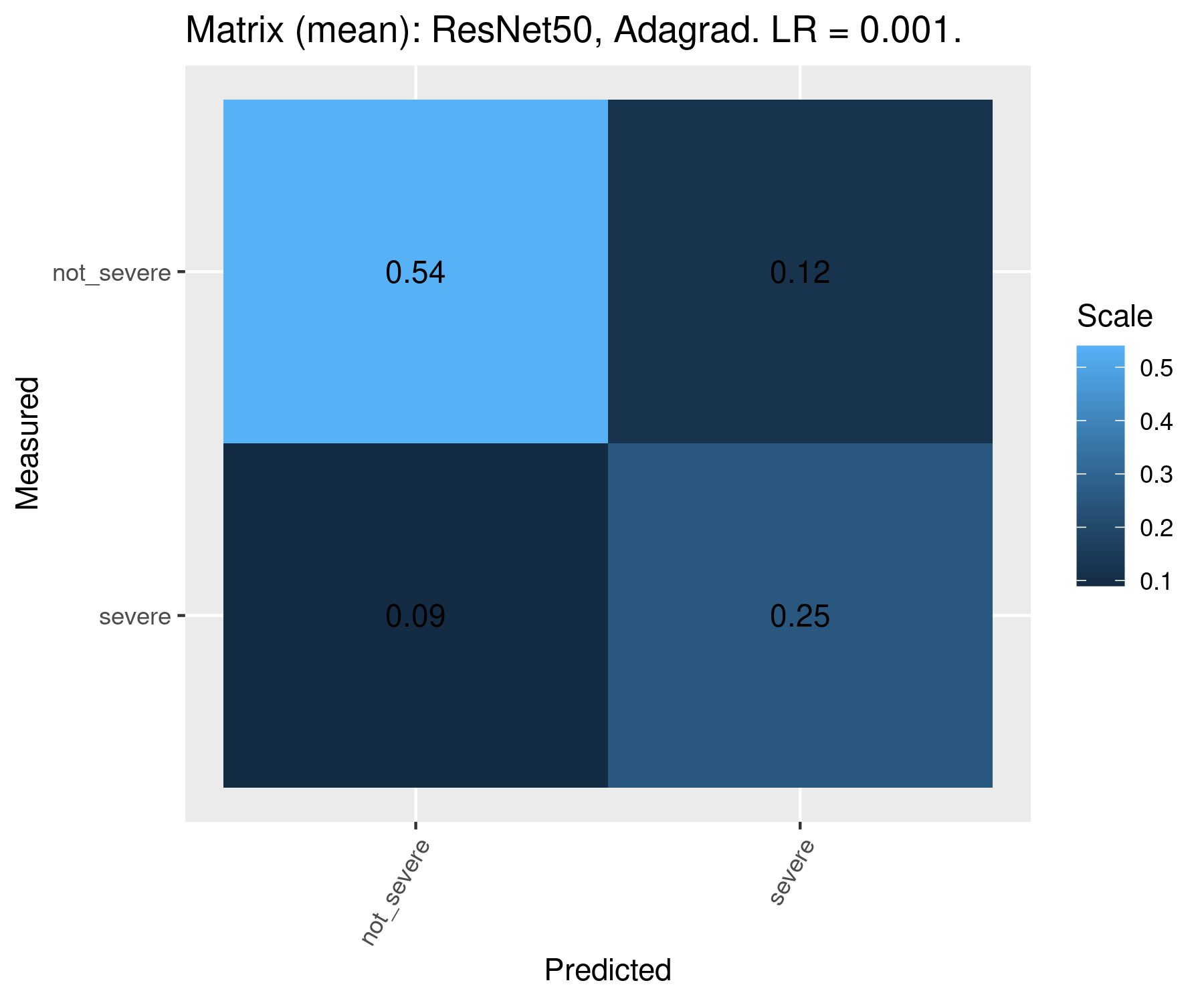}
         \caption{}
         \label{fig:cm_resnet_adagrad_2c}
     \end{subfigure}
     \begin{subfigure}[!t]{0.45\textwidth}
         \centering
         \includegraphics[width=\textwidth]{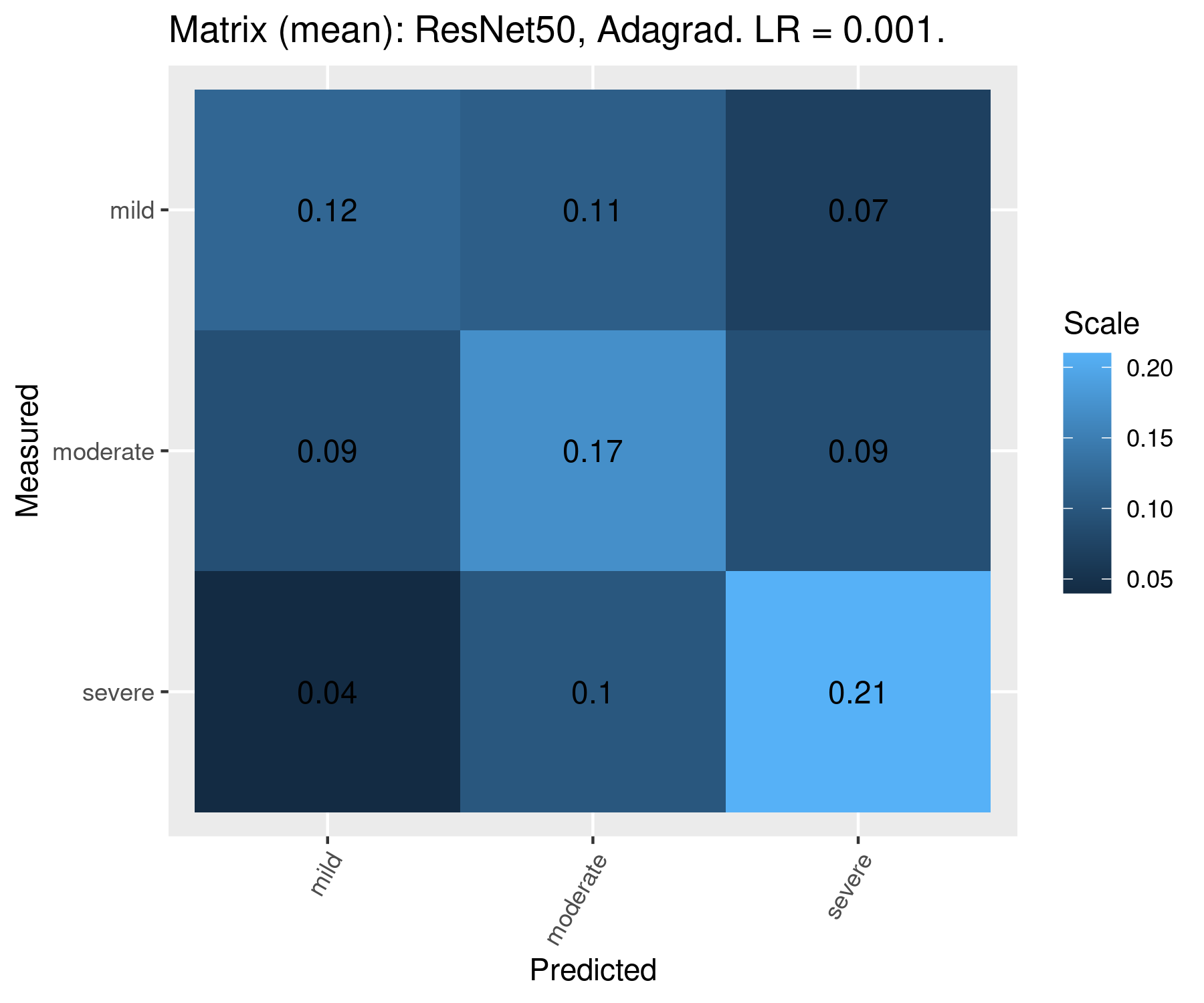}
         \caption{}
         \label{fig:cm_resnet_adagrad_3c}
     \end{subfigure}
        \caption{Confusion matrices for the MobileNetV3 optimized with SAM SGD and for the ResNet50 optimized with Adagrad. The MobileNetV3 achieved the best f-score results for the multiclass classification. ResNet50 achieved the best f-score results for the binary classification.}
        \label{fig:confusion_matrices}
\end{figure*}

Tables~\ref{table:3c_precision},~\ref{table:3c_recall} and~\ref{table:3c_fscore} show statistics for each metric calculated for the analysis. Overall, the best results were achieved by the MobileNetV3 optimized with SAM SGD, and by the ResNet50 optimized with SAM Adagrad. In fact, MobileNetV3 achieved the highest median in all three metrics. The ResNet50 achieved the second best median results in all three metrics when optimized with SAM Adagrad. The recall and f-score results of the MobileNetV3 are also above those of the ResNet50 in that the latter presented outliers whose values were around 30\%, while the MobileNetV3 did not present any of them. Figure~\ref{fig:cm_mobilenet_sam_sgd_3c} shows the normalized confusion matrix for the MobileNetV3. Figure~\ref{fig:cm_resnet_adagrad_3c} shows the normalized confusion matrix for the ResNet50 optimized with Adagrad (not SAM Adagrad), which achieved the best results in the binary classification, provided for comparison. The worst precision and recall results were obtained by the SwinV2 architecture when it was optimized with Adagrad. The worst f-score result was achieved by the MaxViT architecture optimized by SAM SGD.

The two-way ANOVA results for precision suggest the rejection of the null hypothesis for both the architectures and the optimizers: $p=0.0012652$ and $p=0.0235152$, respectively. It also suggests that the interaction between them was relevant ($p=0.0003955$). For recall, the null hypothesis is to be accepted for the optimizers ($p=0.43274$), but the \textit{p-values} for the architectures ($p=0.00097$) and for the interaction ($p=0.00167$) were below the threshold, and hence the null hypothesis is to be rejected for them. For f-score, the ANOVA result was significant for both factors ($p=0.0000225$ and $p=0.0299404$ for architectures and optimizers, respectively) and also for their interaction ($p=0.0000777$). The results of the Scott-Knott clustering test are presented in Tables~\ref{table:3c_precision},~\ref{table:3c_recall} and~\ref{table:3c_fscore}, next to the mean values, where uppercase letters are cluster of optimizers and lowercase letters are clusters of architectures.

\renewcommand{\arraystretch}{1.2}
\begin{table*}
    \caption{Precision statistics for the multiclass problem. The Scott-Knott test results are shown next to the mean values. Uppercase letters in the same rows indicate how optimizers were clustered within the architecture level designated by the row name. Lowercase letters in the same columns indicate how architectures were clustered within the optimizer level designated by the column name.}
    \label{table:3c_precision}
    \centering
    \resizebox{\textwidth}{!}{%
    \begin{tabular}{|c|c|c|c|c|}
    \multicolumn{5}{c}{\textbf{Precision}} \\
    \hline
    \multicolumn{5}{c}{Mean SK (SD)} \\
    \hline
    \multirow{2}{*}{Architecture} & \multicolumn{4}{c|}{Optimizer} \\
    \cline{2-5}
    & SGD & SAM SGD & Adagrad & SAM Adagrad \\
    \cline{1-5}
    ResNet50 & 0.4775142 Aa ($\pm 0.1297515$) & 0.4379582 Aa ($\pm 0.1622991$) & 0.4984670 Aa ($\pm 0.1789589$) & \textbf{0.5743591 Aa ($\pm 0.1371522$)} \\
    MobileNetV3 & 0.4897817 Aa ($\pm 0.1672112$) & 0.5649579 Aa ($\pm 0.1924075$) & 0.4946327 Aa ($\pm 0.1121109$) & 0.4887309 Aa ($\pm 0.1635275$) \\
    DenseNet201 & 0.4905667 Aa ($\pm 0.0971530$) & 0.5118194 Aa ($\pm 0.1197998$) & 0.5110816 Aa ($\pm 0.1969081$) & 0.5140222 Aa ($\pm 0.1921099$) \\
    SwinV2 & 0.5264693 Aa ($\pm 0.1721410$) & 0.1828722 Ba ($\pm 0.1173193$) & 0.2750841 Bb ($\pm 0.1070949$) & 0.5528534 Aa ($\pm 0.1725276$) \\
    MaxViT & 0.5039286 Aa ($\pm 0.1653427$) & 0.1828722 Bb ($\pm 0.1173193$) & 0.4541214 Aa ($\pm 0.1378027$) & 0.3927015 Aa ($\pm 0.1799814$) \\
    \hline
    
    \multicolumn{5}{c}{Median (IQR)} \\
    \hline
    \multirow{2}{*}{Architecture} & \multicolumn{4}{c|}{Optimizer} \\
    \cline{2-5}
    & SGD & SAM SGD & Adagrad & SAM Adagrad \\
    \cline{1-5}
    ResNet50 & 0.4801587 (0.1043015) & 0.4583333 (0.1707912) & 0.4960317 (0.1178922) & 0.5763889 (0.1605159) \\
    MobileNetV3 & 0.5018519 (0.2703310) & \textbf{0.6285354 (0.2931548)} & 0.5000000 (0.0809753) & 0.4583333 (0.1889498) \\
    DenseNet201 & 0.5077381 (0.1034392) & 0.5138889 (0.1806548) & 0.4952381 (0.2921883) & 0.5255342 (0.2480159) \\
    SwinV2 & 0.5629019 (0.1305556) & 0.4409722 (0.4171176) & 0.3195157 (0.1512585) & 0.5777778 (0.2012292)\\
    MaxViT & 0.4347222 (0.2248016) & 0.1176471 (0.1030697) & 0.4317460 (0.1996587) & 0.3951389 (0.1595238)\\
    \hline
    
    \end{tabular}
    }
\end{table*}

\renewcommand{\arraystretch}{1.2}
\begin{table*}
    \caption{Recall statistics for the multiclass problem. The Scott-Knott test results are shown next to the mean values. Uppercase letters in the same rows indicate how optimizers were clustered within the architecture level designated by the row name. Lowercase letters in the same columns indicate how architectures were clustered within the optimizer level designated by the column name.}
    \label{table:3c_recall}
    \centering
    \resizebox{\textwidth}{!}{%
    \begin{tabular}{|c|c|c|c|c|}
    \multicolumn{5}{c}{\textbf{Recall}} \\
    \hline
    \multicolumn{5}{c}{Mean SK (SD)} \\
    \hline
    \multirow{2}{*}{Architecture} & \multicolumn{4}{c|}{Optimizer} \\
    \cline{2-5}
    & SGD & SAM SGD & Adagrad & SAM Adagrad \\
    \cline{1-5}
    ResNet50 & 0.4527201 Aa ($\pm 0.0667939$) & 0.4361616 Ab ($\pm 0.1108422$) & 0.4935137 Aa ($\pm 0.1215467$) & 0.5431313 Aa ($\pm 0.1072030$) \\
    MobileNetV3 & 0.4592280 Aa ($\pm 0.1165084$) & \textbf{0.5558947 Aa ($\pm 0.1343658$)} & 0.4907792 Aa ($\pm 0.0627242$) & 0.4754329 Aa ($\pm 0.1282290$) \\
    DenseNet201 & 0.5218759 Aa ($\pm 0.0510324$) & 0.5116739 Aa ($\pm 0.0751886$) & 0.5145887 Aa ($\pm 0.1287152$) & 0.5123088 Aa ($\pm 0.1425347$) \\
    SwinV2 & 0.4910534 Aa ($\pm 0.1219026$) & 0.4509524 Ab ($\pm 0.1337452$) & 0.3302597 Bb ($\pm 0.0505687$) & 0.4773377 Aa ($\pm 0.1110709$) \\
    MaxViT & 0.4787013 Aa ($\pm 0.0922677$) & 0.3500000 Bc ($\pm 0.0574373$) & 0.4714214 Aa ($\pm 0.0852030$) & 0.4314646 Aa ($\pm 0.0901012$) \\
    \hline
    
    \multicolumn{5}{c}{Median (IQR)} \\
    \hline
    \multirow{2}{*}{Architecture} & \multicolumn{4}{c|}{Optimizer} \\
    \cline{2-5}
    & SGD & SAM SGD & Adagrad & SAM Adagrad \\
    \cline{1-5}
    ResNet50 & 0.4666667 (0.0750000) & 0.4196970 (0.0777778) & 0.4666667 (0.1245491) & 0.5444444 (0.1050505) \\
    MobileNetV3 & 0.4777778 (0.1435967) & \textbf{0.5722222 (0.1444444)} & 0.5166667 (0.0888889) & 0.4777778 (0.1009199) \\
    DenseNet201 & 0.5166667 (0.0583333) & 0.5333333 (0.0638889) & 0.4840548 (0.1333333) & 0.5282107 (0.1722222) \\
    SwinV2 & 0.5052670 (0.1277778) & 0.4388889 (0.2305556) & 0.3166667 (0.0882395) & 0.5000000 (0.0511003) \\
    MaxViT & 0.4879509 (0.1444444) & 0.3333333 (0.0166667) & 0.4666667 (0.1055556) & 0.4239899 (0.0666667) \\
    \hline
    
    \end{tabular}
    }
\end{table*}

\renewcommand{\arraystretch}{1.2}
\begin{table*}
    \caption{F-Score statistics for the multiclass problem. The Scott-Knott test results are shown next to the mean values. Uppercase letters in the same rows indicate how optimizers were clustered within the architecture level designated by the row name. Lowercase letters in the same columns indicate how architectures were clustered within the optimizer level designated by the column name.}
    \label{table:3c_fscore}
    \centering
    \resizebox{\textwidth}{!}{%
    \begin{tabular}{|c|c|c|c|c|}
    \multicolumn{5}{c}{\textbf{F-Score}} \\
    \hline
    \multicolumn{5}{c}{Mean SK (SD)} \\
    \hline
    \multirow{2}{*}{Architecture} & \multicolumn{4}{c|}{Optimizer} \\
    \cline{2-5}
    & SGD & SAM SGD & Adagrad & SAM Adagrad \\
    \cline{1-5}
    ResNet50 & 0.4238737 Ba ($\pm 0.0873516$) & 0.3850030 Bb ($\pm 0.1379018$) & 0.4659213 Aa ($\pm 0.1460202$) & \textbf{0.5363135 Aa ($\pm 0.1107901$)} \\
    MobileNetV3 & 0.4397774 Aa ($\pm 0.1271769$) & 0.5290875 Aa ($\pm 0.1523931$) & 0.4478458 Aa ($\pm 0.0746724$) & 0.4546486 Aa ($\pm 0.1334452$) \\
    DenseNet201 & 0.4836455 Aa ($\pm 0.0772179$) & 0.4741962 Aa ($\pm 0.0683787$) & 0.4804565 Aa ($\pm 0.1406810$) & 0.4897913 Aa ($\pm 0.1492455$) \\
    SwinV2 & 0.4760570 Aa ($\pm 0.1291927$) & 0.3795137 Ab ($\pm 0.1848229$) & 0.2681543 Bb ($\pm 0.0721319$) & 0.4439143 Aa ($\pm 0.1164859$) \\
    MaxViT & 0.4438128 Aa ($\pm 0.0844921$) & 0.2209414 Bc ($\pm 0.0856137$) & 0.4278134 Aa ($\pm 0.1013251$) & 0.3731501 Aa ($\pm 0.1122089$) \\
    \hline
    
    \multicolumn{5}{c}{Median (IQR)} \\
    \hline
    \multirow{2}{*}{Architecture} & \multicolumn{4}{c|}{Optimizer} \\
    \cline{2-5}
    & SGD & SAM SGD & Adagrad & SAM Adagrad \\
    \cline{1-5}
    ResNet50 & 0.4360570 (0.0625291) & 0.3634324 (0.1770259) & 0.4713157 (0.0938957) & 0.5310245 (0.1029412)\\
    MobileNetV3 & 0.4517196 (0.1687585) & \textbf{0.5377345 (0.1877325)} & 0.4652661 (0.0921386) & 0.4634199 (0.1164352) \\
    DenseNet201 & 0.4834355 (0.0645251) & 0.4820110 (0.0878373) & 0.4642941 (0.1745116) & 0.5169553 (0.2149628) \\
    SwinV2 & 0.4981481 (0.1466059) & 0.3619719 (0.3382937) & 0.2689394 (0.0826133) & 0.4693020 (0.1317694) \\
    MaxViT & 0.4228675 (0.1148863) & 0.1739130 (0.0863975) & 0.4112222 (0.1227966) & 0.3878177 (0.0952095) \\
    \hline
    
    \end{tabular}
    }
\end{table*}

\subsection{Two classes}
\label{results_2c}

Tables~\ref{table:2c_precision},~\ref{table:2c_recall} and~\ref{table:2c_fscore} show statistics for the binary classification. It is possible to see that some of the architectures yielded very poor results when optimized with SAM SGD. Table~\ref{table:2c_fscore} shows that ResNet50 optimized with Adagrad achieved both the highest average and median f-score. Although SwinV2 (optimized with SGD and with SAM Adagrad) achieved the highest precision, as one can see in Table~\ref{table:2c_precision}, its recall was very low (Table~\ref{table:2c_recall}, specially when optimized with SAM Adagrad. Figure~\ref{fig:cm_mobilenet_sam_sgd_2c} shows the normalized confusion matrix for the MobileNetV3, the architecture that achieved the best mean and median f-score for multiclass classification, provided for comparison. Figure~\ref{fig:cm_resnet_adagrad_2c} shows the confusion matrix for the ResNet50 optimized with Adagrad, which achieved the best mean and median f-score in the binary classification.

For the binary classification, the two-way ANOVA applied resulted in highly significant p-values for all three metrics. For precision, it yielded $p=5.3 \times 10^{-4}$ for architectures, $p=1.61 \times 10{-11}$ for optimizers, and $p=7.7 \times 10^{-7}$ for their interaction. For recall, p-values were: $p=3.82 \times 10^{-9}$ for architectures, $p=1.0 \times 10^{-11}$ for optimizers, and $p=1.79 \times 10^{-6}$ for their interaction. Finally, p-values for f-score were $p=5.82 \times 10^{-11}$ for architectures, $p=1.16 \times 10^{-16}$ for optimizers, and $p=2.04 \times 10^{-11}$ for their interaction. The results of the Scott-Knott clustering test are presented in Tables~\ref{table:2c_precision},~\ref{table:2c_recall} and~\ref{table:2c_fscore}, next to the mean values, where uppercase letters are cluster of optimizers and lowercase letters are clusters of architectures.

\renewcommand{\arraystretch}{1.2}
\begin{table*}
    \caption{Precision statistics for the binary classification problem. The Scott-Knott test results are shown next to the mean values. Uppercase letters in the same rows indicate how optimizers were clustered within the architecture level designated by the row name. Lowercase letters in the same columns indicate how architectures were clustered within the optimizer level designated by the column name.}
    \label{table:2c_precision}
    \centering
    \resizebox{\textwidth}{!}{%
    \begin{tabular}{|c|c|c|c|c|}
    \multicolumn{5}{c}{\textbf{Precision}} \\
    \hline
    \multicolumn{5}{c}{Mean SK (SD)} \\
    \hline
    \multirow{2}{*}{Architecture} & \multicolumn{4}{c|}{Optimizer} \\
    \cline{2-5}
    & SGD & SAM SGD & Adagrad & SAM Adagrad \\
    \cline{1-5}
    ResNet50 & 0.4144444 Ba ($\pm 0.33423951$) & 0.1500000 Cb ($\pm 0.33747428$) & 0.6796429 Aa ($\pm 0.09637949$) & 0.6025974 Aa ($\pm 0.12504116$) \\
    MobileNetV3 & 0.5096645 Aa ($\pm 0.22255071$) & 0.5274875 Aa ($\pm 0.24217484$) & 0.5899784 Aa ($\pm 0.09019706$) & 0.6064286 Aa ($\pm 0.22433253$) \\
    DenseNet201 & 0.5890842 Aa ($\pm 0.18751978$) & 0.6549908 Aa ($\pm 0.14970299$) & 0.6561905 Aa ($\pm 0.17821185$) & 0.6707035 Aa ($\pm 0.13704775$) \\
    SwinV2 & \textbf{0.7213095 Aa ($\pm 0.20103227$)} & 0.0000000 Cb ($\pm 0.00000000$) & 0.3500000 Bb ($\pm 0.30882094$) & 0.6214286 Aa ($\pm 0.45928318$) \\
    MaxViT & 0.5571429 Aa ($\pm 0.31430976$) & 0.1333333 Bb ($\pm 0.32203059$) & 0.6485714 Aa ($\pm 0.14283598$) & 0.6685684 Aa ($\pm 0.19827464$) \\
    \hline
    
    \multicolumn{5}{c}{Median (IQR)} \\
    \hline
    \multirow{2}{*}{Architecture} & \multicolumn{4}{c|}{Optimizer} \\
    \cline{2-5}
    & SGD & SAM SGD & Adagrad & SAM Adagrad \\
    \cline{1-5}
    ResNet50 & 0.4722222 (0.51666667) & 0.0000000 (0.00000000) & 0.6458333 (0.14107143) & 0.5982143 (0.14236111) \\
    MobileNetV3 & 0.4642857 (0.17654221) & 0.5701754 (0.29523810) & 0.6000000 (0.13750000) & 0.5833333 (0.34107143) \\
    DenseNet201 & 0.5549451 (0.16666667) & 0.6201923 (0.15773810) & 0.6666667 (0.23863636) & 0.6458333 (0.11212121) \\
    SwinV2 & 0.7500000 (0.20476190) & 0.0000000 (0.00000000) & 0.5000000 (0.62500000) & \textbf{0.8571429 (0.87500000)} \\
    MaxViT & 0.5000000 (0.39285714) & 0.0000000 (0.00000000) & 0.6333333 (0.08809524) & 0.5902778 (0.25705128) \\
    \hline
    
    \end{tabular}
    }
\end{table*}

\renewcommand{\arraystretch}{1.2}
\begin{table*}
    \caption{Recall statistics for the binary classification problem. The Scott-Knott test results are shown next to the mean values. Uppercase letters in the same rows indicate how optimizers were clustered within the architecture level designated by the row name. Lowercase letters in the same columns indicate how architectures were clustered within the optimizer level designated by the column name.}
    \label{table:2c_recall}
    \centering
    \resizebox{\textwidth}{!}{%
    \begin{tabular}{|c|c|c|c|c|}
    \multicolumn{5}{c}{\textbf{Recall}} \\
    \hline
    \multicolumn{5}{c}{Mean SK (SD)} \\
    \hline
    \multirow{2}{*}{Architecture} & \multicolumn{4}{c|}{Optimizer} \\
    \cline{2-5}
    & SGD & SAM SGD & Adagrad & SAM Adagrad \\
    \cline{1-5}
    ResNet50 & 0.30454545 Bb ($\pm 0.26247568$) & 0.03333333 Cb ($\pm 0.07027284$) & \textbf{0.73030303 Aa ($\pm 0.23943869$)} & 0.67272727 Aa ($\pm 0.20875149$) \\
    MobileNetV3 & 0.50454545 Aa ($\pm 0.19298614$) & 0.51515152 Aa ($\pm 0.25191846$) & 0.65000000 Aa ($\pm 0.21444732$) & 0.68939394 Aa ($\pm 0.26612376$) \\
    DenseNet201 & 0.59696970 Aa ($\pm 0.28482341$) & 0.58939394 Aa ($\pm 0.21504123$) & 0.67272727 Aa ($\pm 0.29459644$) & 0.62272727 Aa ($\pm 0.19592445$) \\
    SwinV2 & 0.57121212 Aa ($\pm 0.22394306$) & 0.00000000 Cb ($\pm 0.00000000$) & 0.26969697 Bb ($\pm 0.29697657$) & 0.29242424 Bb ($\pm 0.34477235$) \\
    MaxViT & 0.40000000 Bb ($\pm 0.25092422$) & 0.11666667 Cb ($\pm 0.31476034$) & 0.60151515 Aa ($\pm 0.23527440$) & 0.64696970 Aa ($\pm 0.19948536$) \\
    \hline
    
    \multicolumn{5}{c}{Median (IQR)} \\
    \hline
    \multirow{2}{*}{Architecture} & \multicolumn{4}{c|}{Optimizer} \\
    \cline{2-5}
    & SGD & SAM SGD & Adagrad & SAM Adagrad \\
    \cline{1-5}
    ResNet50 & 0.3333333 (0.4583333) & 0.0000000 (0.0000000) & 0.7500000 (0.4242424) & \textbf{0.7803030 (0.2916667)} \\
    MobileNetV3 & 0.5000000 (0.2613636) & 0.5000000 (0.2916667) & 0.5833333 (0.3333333) & 0.6969697 (0.1666667) \\
    DenseNet201 & 0.6666667 (0.3825758) & 0.6666667 (0.2121212) & 0.6969697 (0.4583333) & 0.5833333 (0.2121212) \\
    SwinV2 & 0.6060606 (0.1666667) & 0.0000000 (0.0000000) & 0.1666667 (0.5909091) & 0.1666667 (0.5189394) \\
    MaxViT & 0.4166667 (0.2916667) & 0.0000000 (0.0000000) & 0.6666667 (0.1666667) & 0.6515152 (0.2916667) \\
    \hline
    
    \end{tabular}
    }
\end{table*}

\renewcommand{\arraystretch}{1.2}
\begin{table*}
    \caption{F-Score statistics for the binary classification problem. The Scott-Knott test results are shown next to the mean values. Uppercase letters in the same rows indicate how optimizers were clustered within the architecture level designated by the row name. Lowercase letters in the same columns indicate how architectures were clustered within the optimizer level designated by the column name.}
    \label{table:2c_fscore}
    \centering
    \resizebox{\textwidth}{!}{%
    \begin{tabular}{|c|c|c|c|c|}
    \multicolumn{5}{c}{\textbf{F-Score}} \\
    \hline
    \multicolumn{5}{c}{Mean SK (SD)} \\
    \hline
    \multirow{2}{*}{Architecture} & \multicolumn{4}{c|}{Optimizer} \\
    \cline{2-5}
    & SGD & SAM SGD & Adagrad & SAM Adagrad \\
    \cline{1-5}
    ResNet50 & 0.32321789 Bb ($\pm 0.2529390$) & 0.05357143 Cb ($\pm 0.1132518$) & \textbf{0.68324608 Aa ($\pm 0.1354936$)} & 0.61345845 Aa ($\pm 0.1306311$) \\
    MobileNetV3 & 0.48682984 Aa ($\pm 0.1592728$) & 0.50613609 Aa ($\pm 0.2235576$) & 0.60394046 Aa ($\pm 0.1214086$) & 0.60941466 Aa ($\pm 0.1929006$) \\
    DenseNet201 & 0.54167582 Aa ($\pm 0.1862695$) & 0.59311661 Aa ($\pm 0.1508597$) & 0.61857239 Aa ($\pm 0.1798277$) & 0.62811855 Aa ($\pm 0.1137390$) \\
    SwinV2 & 0.60800804 Aa ($\pm 0.1747624$) & 0.00000000 Cb ($\pm 0.0000000$) & 0.28753501 Bb ($\pm 0.2857950$) & 0.32180403 Bb ($\pm 0.3090197$) \\
    MaxViT & 0.44542735 Bb ($\pm 0.2516300$) & 0.07857143 Cb ($\pm 0.1731723$) & 0.59483239 Aa ($\pm 0.1606645$) & 0.62963203 Aa ($\pm 0.1178070$) \\
    \hline
    
    \multicolumn{5}{c}{Median (IQR)} \\
    \hline
    \multirow{2}{*}{Architecture} & \multicolumn{4}{c|}{Optimizer} \\
    \cline{2-5}
    & SGD & SAM SGD & Adagrad & SAM Adagrad \\
    \cline{1-5}
    ResNet50 & 0.3928571 (0.4868687) & 0.0000000 (0.0000000) & \textbf{0.7207792 (0.1478261)} & 0.6200000 (0.1553221) \\
    MobileNetV3 & 0.4807692 (0.1281219) & 0.5307692 (0.2174603) & 0.5668449 (0.1534702) & 0.6599327 (0.2448524) \\
    DenseNet201 & 0.5773810 (0.1642628) & 0.6153846 (0.0562500) & 0.6478261 (0.1998663) & 0.6410256 (0.1432900) \\
    SwinV2 & 0.6234818 (0.2022727) & 0.0000000 (0.0000000) & 0.2500000 (0.5462185) & 0.2857143 (0.4985119)  \\
    MaxViT & 0.4000000 (0.3663462) & 0.0000000 (0.0000000) & 0.5934066 (0.2178030) & 0.6666667 (0.1279762) \\
    \hline
    
    \end{tabular}
    }
\end{table*}

%%%%%%%%%%%%%%  D I S C U S S Ã O %%%%%%%%%%%%%%%%%%%%%%

\section{Discussion}

First of all, Table~\ref{table:num_images} shows that the number of images of non stenotic nares in the dataset is limited to three. This is due to the fact that non stenotic brachycephalic dogs are rare. From the same Table~\ref{table:num_images}, one can see that the dataset is not heavily imbalanced when used for multi-class classification. It is imbalanced, though, when used for binary classification, with two non severely stenotic images for each severely stenotic one. This means that Figure~\ref{fig:confusion_matrices} can be read straightforwardly for the multi-class classification, but one should aware of the imbalance when looking at the confusion matrices for binary classification.

As stated in Section~\ref{results}, some of the networks yielded very poor results when optimized by SAM SGD. Further inspection of the results shows that in these cases the neural networks ``learned'' to classify all images as non severe in most of the folds (in fact, SwinV2 did so in all ten folds). This is likely a problem with the optimization, which is to be expected, since the hyperparameters were kept the same for a large number of model/optimizer combinations. Data augmentation hyperparameters and optimizer hyperparameters were not changed during the experiment as well, both of which can hugely impact the performance. This is a possible improvement for further experiments.

One can argue from Figure~\ref{fig:cm_mobilenet_sam_sgd_3c} that the severe class was the one more easily identified by the MobileNetV3/SAM SGD. It is also possible to claim that the network biased, albeit not so much, towards saying that the situation was worse than it actually was, since almost half of the predictions made by the model were of severe stenosis (\textit{i.e.}, approximately 44\%, against 34\% for mild and 23\% for moderate stenosis). The middle row in Figure~\ref{fig:cm_mobilenet_sam_sgd_3c} also shows that the moderate stenosis was the most difficult for the network to learn, since approximately two thirds of the images actually belonging to the moderate class were wrongly classified, either as mild or as severe.

A comparison between Figures~\ref{fig:cm_mobilenet_sam_sgd_2c} and~\ref{fig:cm_mobilenet_sam_sgd_3c} suggests that modelling the problem as one of binary classification is not helpful for the MobileNetV3/SAM SGD. This may be relevant for the purposes of an end-to-end smartphone system, since correct predictions of severe stenosis are more desirable, even if precision is sacrificed. On a related note, this is also the reason why the SwinV2 results for binary classification are less good then they may seem from Table~\ref{table:2c_precision} alone. Given the high precision with low recall combination, one can speculate that changes to the classification threshold may prove beneficial for the performance of this architecture, but one should still keep in mind that the SwinV2 architecture is the largest one evaluated in this work, with almost 88 million parameters, and it is unlikely that it differs from the much smaller ResNet50/Adagrad.

Figures~\ref{fig:cm_mobilenet_sam_sgd_2c} and~\ref{fig:cm_resnet_adagrad_3c} show that the behavior of the ResNet50 optimized with Adagrad was quite different than that of MobileNetV3/SAM SGD regarding the change in the number of classes. Changing the problem into one of binary classification seems to have improved the capacity of the model to identify severe stenosis, without sacrificing its capacity to identify non severe stenosis. Arguably, this is also related to the fact that the ResNet50/Adagrad did not have as much difficulty identifying moderate stenosis as did the MobileNetV3/SAM SGD. Overall, the ResNet50/Adagrad seems to be better at identifying severe stenosis, but the high standard deviation shown in Table~\ref{table:2c_recall} and the ANOVA results show that it is ultimately not certain that the models' performance differ, and further considerations are necessary regarding the advantages of using a heavier ResNet50 instead of a MobileNetV3.

\begin{figure*}
     \centering
     \begin{subfigure}[b]{0.3\textwidth}
         \centering
         \includegraphics[width=\textwidth,height=\textwidth]{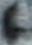}
         \caption{}
         \label{fig:severe_hard_03}
     \end{subfigure}
     \hfill
     \begin{subfigure}[b]{0.3\textwidth}
         \centering
         \includegraphics[width=\textwidth,height=\textwidth]{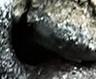}
         \caption{}
         \label{fig:severe_hard_09}
     \end{subfigure}
     \hfill
     \begin{subfigure}[b]{0.3\textwidth}
         \centering
         \includegraphics[width=\textwidth,height=\textwidth]{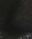}
         \caption{}
         \label{fig:severe_hard_030}
     \end{subfigure}
        \caption{Hard cases of severe stenosis images. Neither the MobileNetV3/SAM SGD nor the ResNet50/Adagrad classified them correctly in the multi-class classification and in the binary classification problems, respectively.}
        \label{fig:severe_hard_case}
\end{figure*}

Figure~\ref{fig:severe_hard_case} shows examples of images of severe stenosis that were incorrectly classified by the best architectures (by median f-score). One can consider that these are hard cases. If, as stated, stenosis is visually characterized by a tissue that covers too much of the nostril, then it is possible to see why Figure~\ref{fig:severe_hard_09} would be incorrectly classified. Since in some cases stenosis is also identified in virtue of other symptoms, then the challenge in classifying them using images alone rises. Figures~\ref{fig:severe_hard_03} and~\ref{fig:severe_hard_030} shows other difficulties presented by the task. Images can have low resolutions, or be taken in such a way that even the nostril is not adequately visible.

%Parágrafo 9: Criar possíveis explicações para os números relevantes que apareceram nos resultados. Argumentar motivas de alguma técnicas ter se saído melhor ou pior que outras.

%Parágrafo 10: Quando possível, comparar os resultados com os de alguns outros artigos que tratam problemas similares.

%Parágrafo 11: Mostrar vários tipos de erros cometidos pelas técnicas e descobrir possíveis motivos para eles terem ocorrido. Destacar características do banco de imagens que o tornam mais desafiador.

%Parágrafo 12: Discutir deficiências do experimento que podem ser aprimoradas no futuro e possíveis formas de melhorar o desempenho que não puderam ainda ser testadas.

%%%%%%%%%%%%%%  C O N C L U S O E S %%%%%%%%%%%%%%%%%%%%%%%%%%%%%%%%%%%%%%%%%%%%%

\section{Conclusion}

In this work, we presented the first image dataset for classifying stenosis degree in brachycephallic dog breeds. The dataset is publicly available. We also presented the evaluation of several neural networks, achieving a maximum average recall of 73\% and a maximum average f-score of 68\% for the identification of severe stenosis. The results indicate that, although the problem is a hard one, a solution for it may be possible. Arguably, properly solving it may assist veterinarians who have to deal with stenosis cases and also tutors, who do not have the proper training to identify severe stenosis. Finally, it may relieve many of the animals from their suffering.

\section{Acknowledgments}
This work has received financial support from the Dom Bosco Catholic University and the Foundation for the Support and Development of Education, Science and Technology from the State of Mato Grosso do Sul, FUNDECT. Some of the authors have been awarded with Scholarships from the the Brazilian National Council of Technological and Scientific Development, CNPq and the Coordination for the Improvement of Higher Education Personnel, CAPES.

% ==============================================================

% \appendix

% \section{Section in Appendix}
% \label{appendix-sec1}

%% References

% \bibliographystyle{elsarticle-num}
%\bibliographystyle{elsarticle-harvard}

%\bibliography{references}

\end{document}